\documentclass[a4paper]{jpconf}
\usepackage{graphicx}

\begin{document}
\title{QPO in RE J1034+396: model constraints from observed trends}

\author{Bozena Czerny, Pawel Lachowicz}
\address{Copernicus Astronomical Center, Bartycka 18, 00-716 Warsaw, Poland}
\ead{bcz@camk.edu.pl}
\author{Michal Dov\v{c}iak, Vladim\'{\i}r Karas, Tom\'a\v{s} Pech\'a\v{c}ek}
\address{Astronomical Institute, Academy of Sciences, Bo\v{c}n\'{\i} II 1401, 14131 Prague, Czech Republic}
\author{Tapas K. Das}
\address{Harish Chandra Research Institute, Jhunsi, Allahabad 211 019, India}

\begin{abstract}
We analyze the time variability of the X-ray emission of RE J1034+396, an active galactic 
nucleus with the first firm detection of a quasi-periodic oscillations (QPO). Based on the 
results of a wavelet analysis, we find a drift in the QPO central frequency. The change in
the QPO frequency correlates with the change in the X-ray flux with a short time delay. 
Linear structures such as shocks, spiral waves, or
very distant flares seem to be a favored explanation for this particular QPO event.
\end{abstract}

\section{Introduction}

The QPO appearance has been reported to occur in a bright Seyfert 1 galaxy, RE
J1034+396 (Gierlinski et al. 2008) and BL Lacertae object PKS 2155-304 (Lachowicz
et al. 2009). So far the QPO origin has not been well understood. The
wavelet analysis can be particularly useful in the exploration of the QPO properties.
Similar to Fourier spectra methods, the wavelet analysis provides fast linear operations
on data vectors. Unlike the Fourier transform, the base functions of wavelets
are localized, and this brings new opportunities to the data analysis. Based on the
wavelet results we discuss possible constraints on the physical origin of the QPO in
RE J1034+396.

\begin{figure}[t]
\begin{minipage}{.46\linewidth}

\includegraphics[width=15pc]{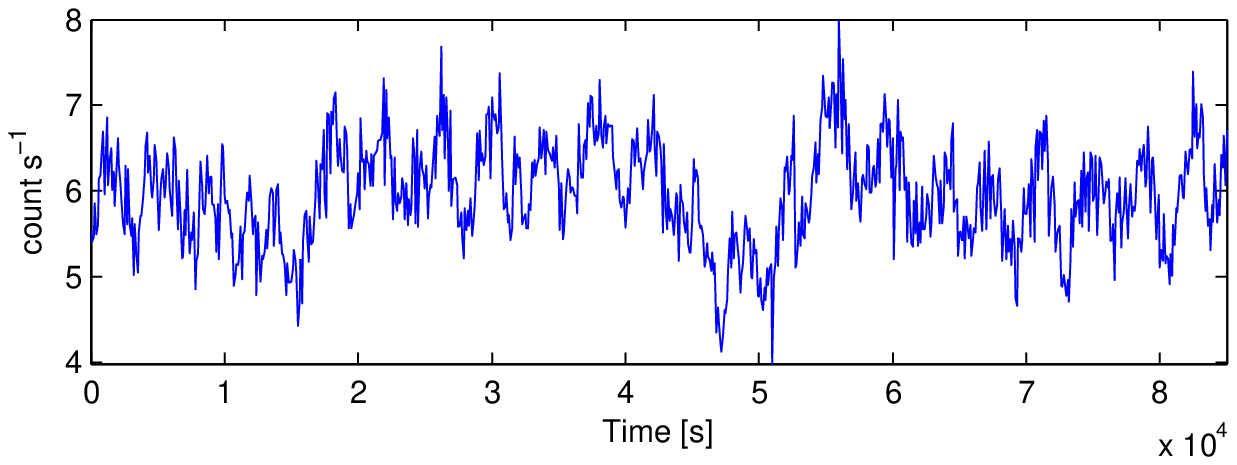}
\caption{\label{fig1a}
The X-ray light curve of RE J1034+396. The significant regions are within a period range
of 3000--4500 s, where they form a rather complex pattern. The corresponding wavelet
map is shown in Figure~2.}
\end{minipage} \hspace{2pc}%
\begin{minipage}{.46\linewidth}
\vspace*{-24mm}\includegraphics[width=15pc]{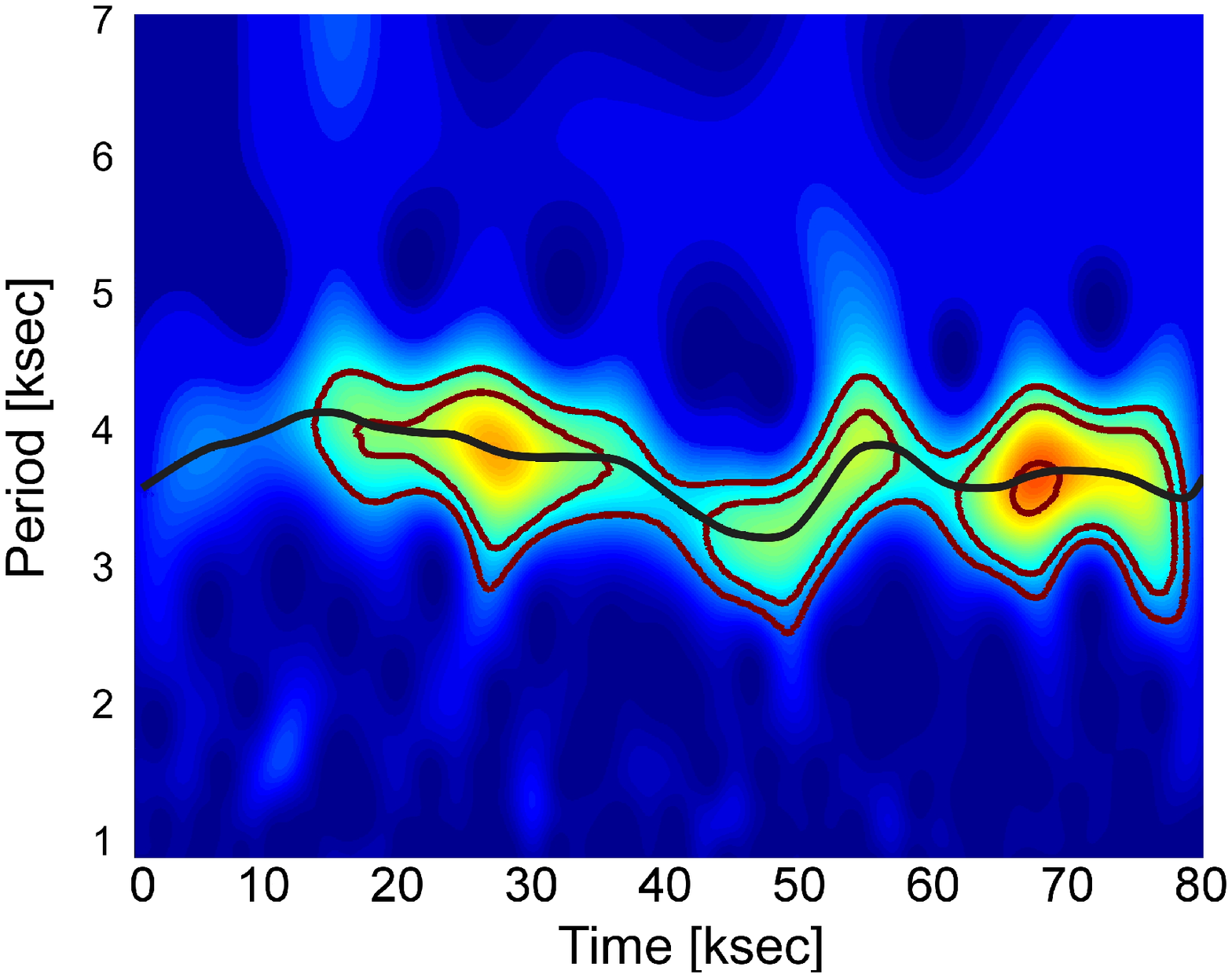}
\caption{\label{fig1b}
The increasing values of wavelet power are denoted by gradual changes of the
colours, going from deep blue to red. The confidence contours
of 90\%, 95\%, and 99\% are marked. A solid black line traces the the QPO period.}
\end{minipage} 
\end{figure} 

\section{Wavelet analysis of RE J1034+396 light curve}

The source RE J1034+396 was observed by XMM-Newton satellite on 2007 May 31,
and the resulting X-ray light curve analysis was reported in G08. Using the same
procedure, we extract the source light curve in the 0.3-10 keV energy band, which
covers $8.5 \times 10^4$ s of continuous observations sampled evenly every $t=100$ s 
(see Figure~\ref{fig1a}). We
use the wavelet analysis codes of Torrence \& Compo (1998) to analyse the source behaviour.

The wavelet transform can be compared with the Fourier transform, which is
used routinely to represent a measured light curve as a superposition of
sine and cosine components. Both transforms can be classified into discrete
versus continuous forms. In certain situations, wavelets can provide a more
suitable representation of the signal by allowing a balanced resolution
between time and frequency domains. Furthermore, the discrete wavelet
transform can be less computationally expensive, as it requires ${\cal
O}(N)$ time steps compared to ${\cal O}(N \log N)$ steps for the
corresponding resolution in fast Fourier transform.

The wavelet map confirms the presence of the QPO in the source at the period $P_0 \sim
3733$ s through a range of peaks. In order to assign the confidence levels to those
peaks we have performed Monte Carlo simulations (Czerny et al. 2010). We created
7500 simulated lightcurves and the corresponding wavelet maps. Then we built the
statistics for each of the scales in the discussed range. This allowed us to assign
significance levels for each wavelet scale independently, which is important in the
lightcurves with underlying red noise background. The average wavelet values rise
toward longer periods, as does the power density spectrum, and this drift causes the
observed misalignment between the colors on the map (referring to absolute values)
and confidence levels in Figure~\ref{fig1b}.
In order to find whether any additional periodicity is present in the system, we analyzed
a temporal variability of the QPO period. The periodogram analysis indicates
a time-scale of $\sim24$ ks but the formal significance
of the periodicity is low because of the high red-noise level.

\section{Flux versus QPO period trend}

\begin{figure}[t]
\begin{minipage}{.46\linewidth}

\includegraphics[width=17pc]{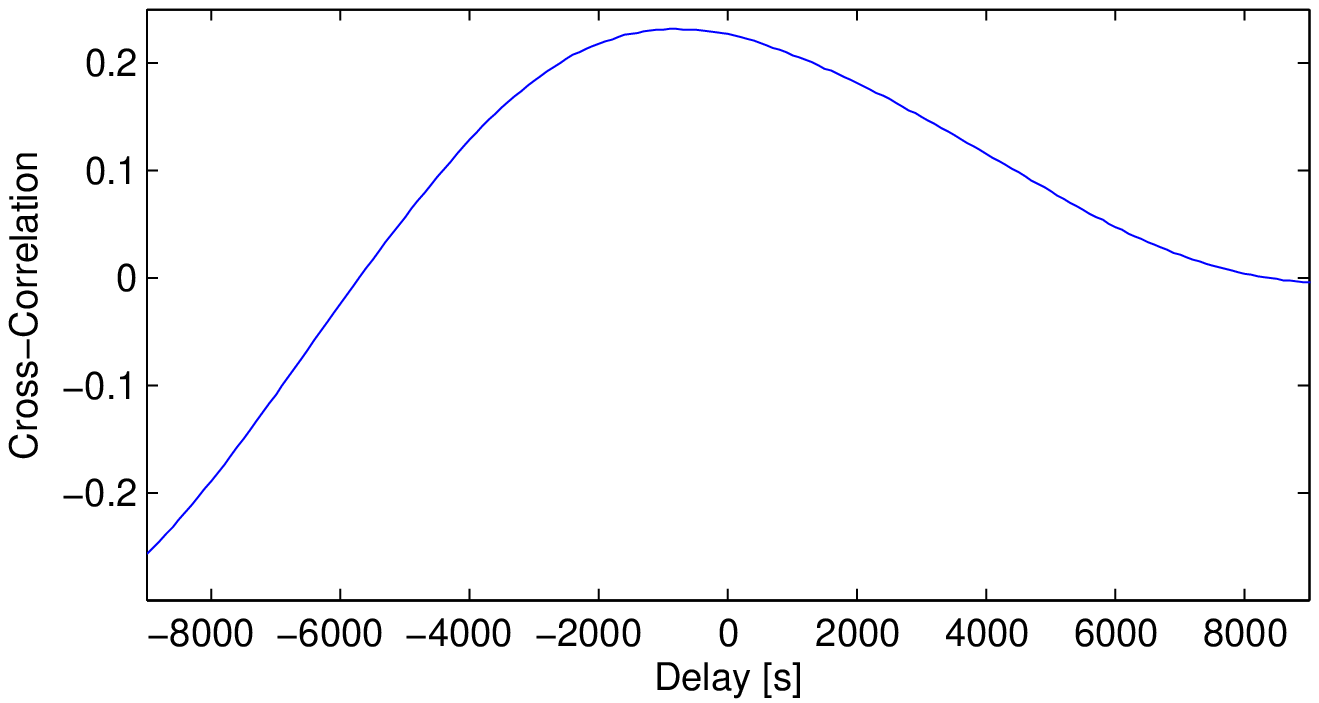}
\caption{\label{cross}
The normalized cross-correlation function of the QPO period and the X-ray light curve 
in the Segment 2.}
\end{minipage} \hspace{1pc}%
\begin{minipage}{.46\linewidth}
\vspace*{-9.pc}
\includegraphics[width=20.5pc]{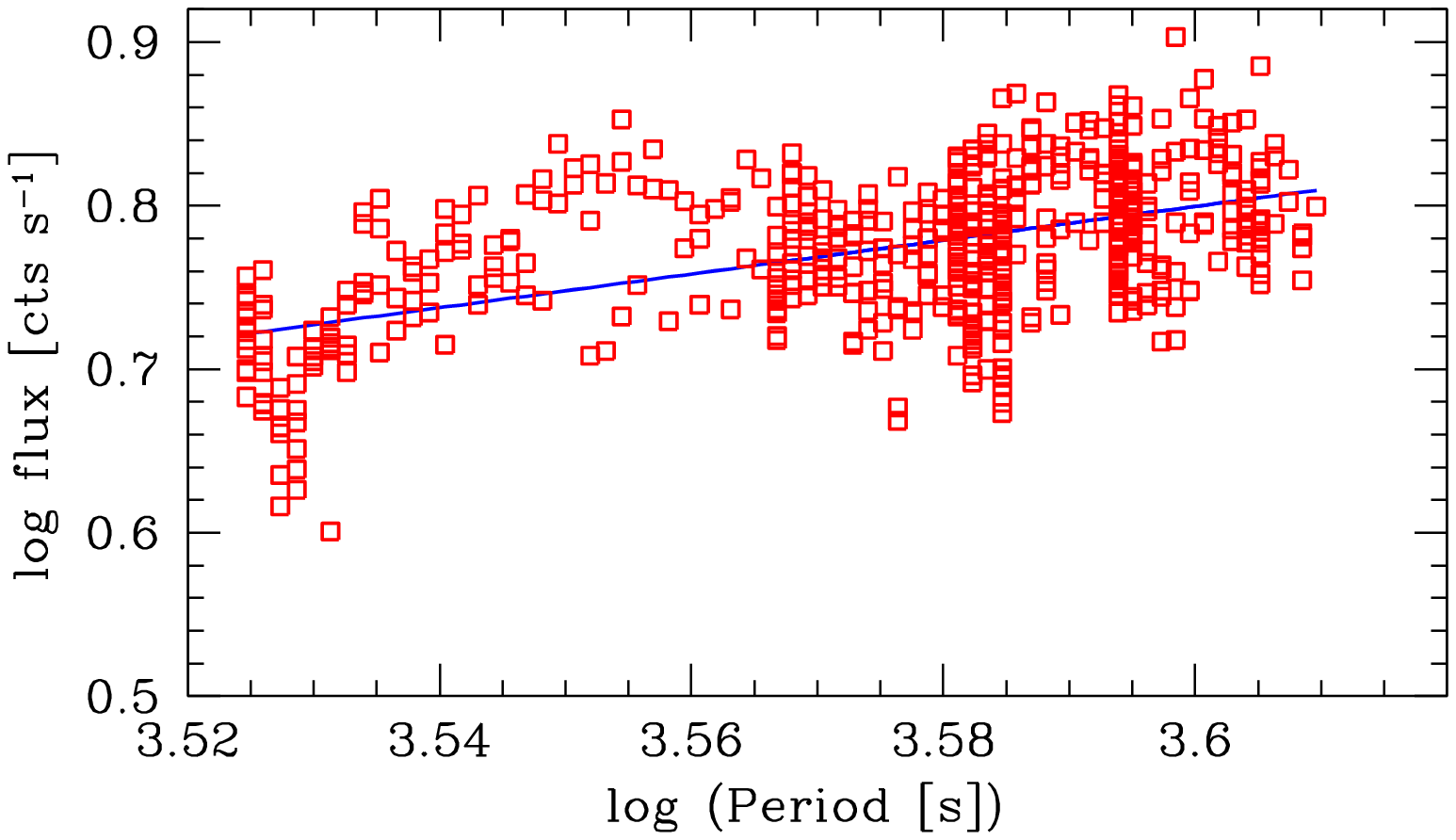}
\vspace*{-3.5pc}\caption{\label{flux_period}The flux versus QPO period relation in the same Segment 2 
of the data as in Figure 3.}
\end{minipage} 
\end{figure}

The wavelet map shows a strong decrease in the marked QPO period at the time about
$5 \times 10^4$ s, which is accompanied by a decrease in the X-ray flux. This motivated us
to check for the overall correlation between the flux and the QPO period in the data.
The correlation is present, and the time delay between the curves
is ~900 s, with the flux lagging behind the frequency change. The delay is by a
factor 4.1 shorter than the QPO period itself. If only the second part of the data is
investigated (Segment 2 in notation of G08), the measured delay is somewhat longer, 1400 s.
The cross-correlation function between the X-ray luminosity and the QPO period for Segment 2 is
shown in Figure~\ref{cross}, and the corresponding Flux-Period relation is given in 
Figure~\ref{flux_period}. 
It suggests that some underlying mechanism must be responsible
for observed modulation of the QPO amplitude, the frequency and the flux.

Since the QPO of a 3733 s period smears the correlation at the longer time-scales,
we extract the short-time trend from the light curve by applying the moving average
approach, with a 5000 s window function. In consequence, we obtain a smoothed
curve, free from temporal high-frequency flux variability. The scatter is the lower,
as expected, but the overall trend is the same (see Czerny et al. 2010).
Because the amplitude of the wavelet map also varies considerably, we tested for
other possible correlations as well. The weak correlation between the wavelet amplitude
and the QPO period was present but it was less significant than the correlation
of X-ray flux vs. QPO period.

\section{Shock oscillation scenario}

\begin{figure}[t]
\begin{minipage}{.46\linewidth}

\includegraphics[width=19pc]{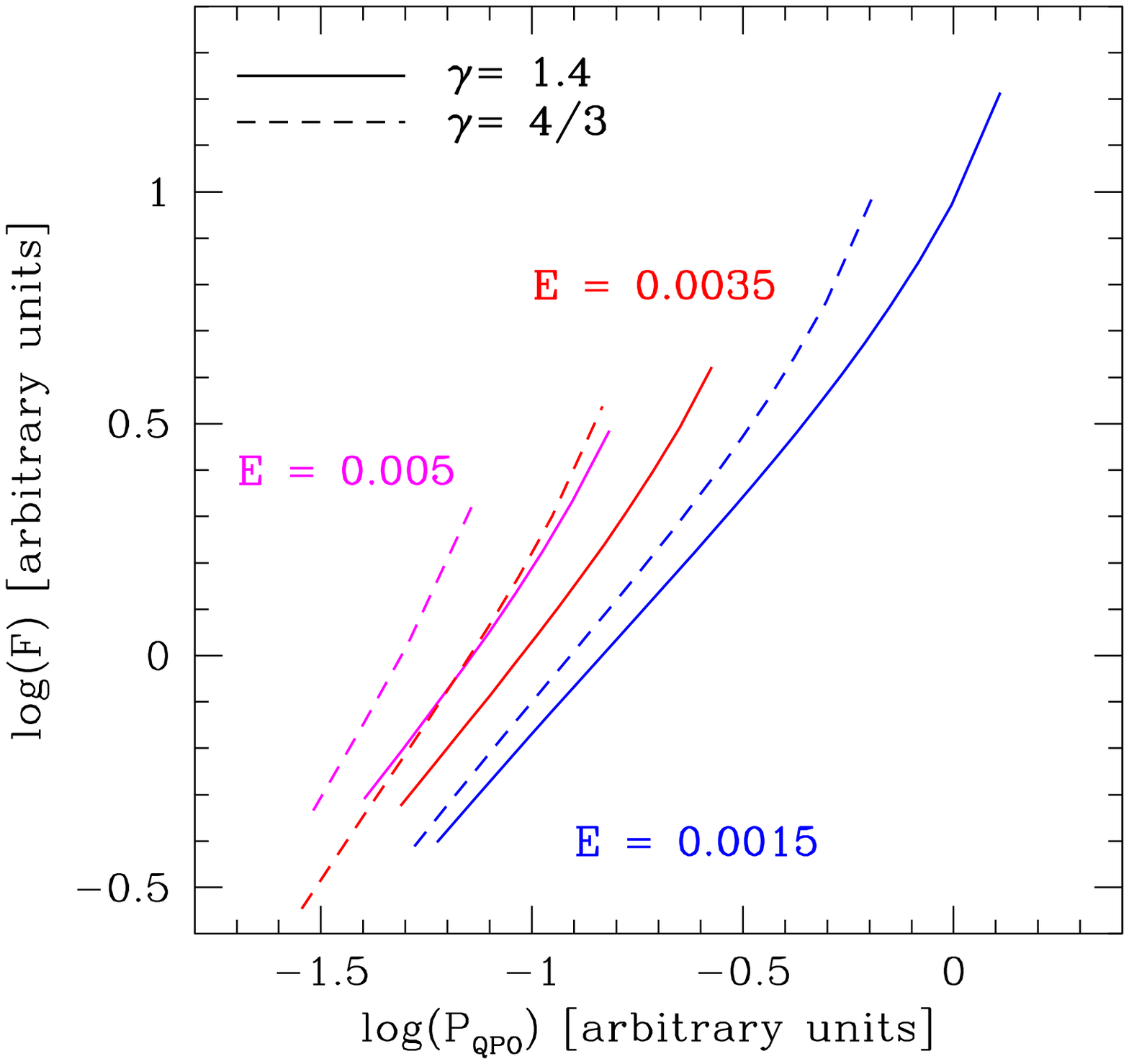}
\caption{\label{fig43}
Flux vs. period relation for
a fixed value of specific energy ${\cal E}$, and variable angular 
momentum, $\lambda$, for three pairs of solutions for three different values of
${\cal E}$. Each pair is characterized by two different values of the 
adiabatic indices, $\gamma
 = 4/3$ (dashed curve) and 
 $\gamma = 1.4$
(solid curve).}
\end{minipage} \hspace{2pc}%
\begin{minipage}{.46\linewidth}
\vspace*{-5mm}
\includegraphics[width=19pc]{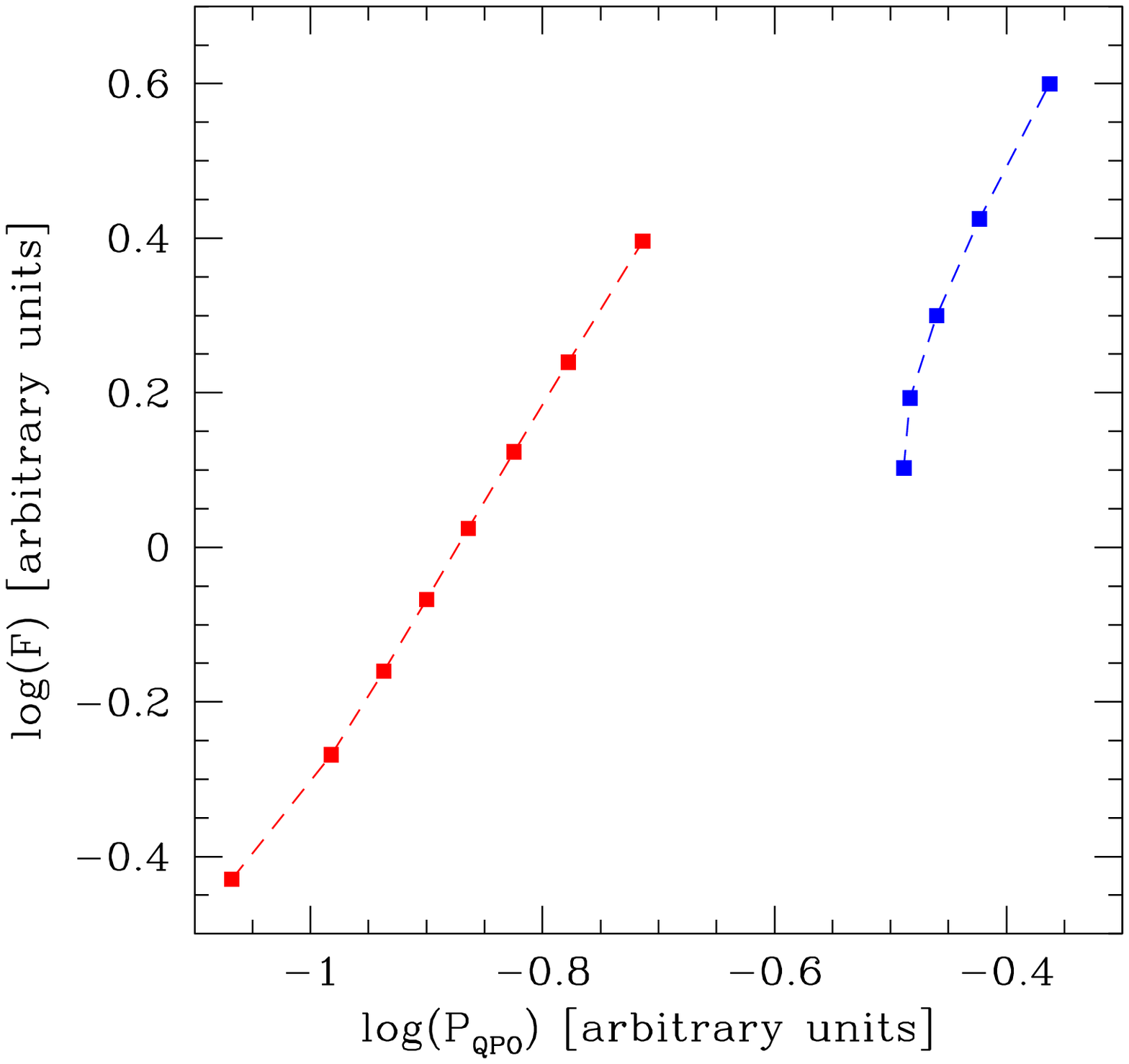}
\caption{\label{fig1725}
Flux vs.\ period relation for
a fixed angular momentum, $\lambda$, and variable asymptotic value of
the Bernoulli constant, ${\cal E}$. Two sequences represent two values 
of $\lambda$: 1.725 (left curve)
and 1.775 (right curve), $\gamma = 4/3$.}
\end{minipage} 
\end{figure}

We now study in detail one specific scenario of the QPO formation. In this model, QPO 
oscillations are due to the oscillations of the shock formed in the low angular momentum
hot accretion flow, and the variation of the shock location corresponds to the observed
changes in the QPO period and the X-ray flux. We calculate the shock location as a function
of parameters for a stationary flow in the Kerr black hole metric. We calculate the
QPO frequency from the Keplerian frequency at the shock location and the compression term, since
the oscillations are  not strictly dynamical, as shown by Molteni (1996). The X-ray flux
attributed to the shock is estimated from the thermal energy flux across the shock.
Next, we consider the variations of the flow parameters in quasi-stationary approach and
study the relation between the flux and the period when either angular momentum of
the matter or the Bernoulli constant vary. The examples of the Flux-Period coupling are shown
in Figures~\ref{fig43} and ~\ref{fig1725}.

We obtain that a change in the shock
location caused by perturbation of the flow angular momentum is compatible with
the trends observed in RE J1034+396, whereas the perturbation of the specific flow
energy results in too strong flux response to the change of the oscillation period. 
Finally, we examined whether the results depend significantly on the spin of the black hole. 
However, since the shock forms not very close to the black hole, the effect is not strong.
The slope of the Flux-Period relation is practically unchanged, the shock frequency itself
varies with the Kerr parameter but this information cannot be used to determine the spin
unless the black hole mass is known. 

\section{Discussion and Conclusions}

Our analysis of the X-ray light curve of RE J1034+396 through the wavelet technique
suggests the dependence of the QPO peak position in a single event on time, and the change of the
QPO period correlates with the X-ray flux. This correlation implies a common source
of perturbation. If the result
indicating the proportionality between the flux and the QPO period is a more general
property, then it implies strong constraints on the QPO mechanism. In particular, it
appears to contradict the explanation in terms of an orbiting hot spot model and it 
supports the linear structures with intrinsic timescales like shocks, or spiral waves.

 In Galactic sources,
single QPO high-frequency events are currently unresolved, therefore this newly found correlation
cannot be tested. However, future X-ray observations should bring more light
curves of AGN with a quality good enough to perform a similar analysis, even if the
QPO duty cycle is relatively low (Middleton et al. 2011).

There are also other approaches than the wavelet analysis which allow for a deeper insight into
the QPO phenomenon. Zoghbi \& Fabian (2011) examined time-lags of the signal arriving at different X-ray
energy bands. In this way they reveal different reverberation properties of
the source during the period when the QPO appeared and when it was not
present.  Middleton et al. (2011) studied the 
covariance spectra and identified the soft component accompanying the QPO hotter than the soft X-ray excess
seen in the average spectra. Both results imply that QPO signal is
modulated at a distance of a few to few tenth of gravitational radii from the black hole. 
 
\medskip

This work has been supported in part by NN203 380136 grant in Poland, the XIth plan at HRI in India,
and PECS 98040 in Czech Republic.



\end{document}